\documentclass[fleqn,useAMS,usenatbib]{mnras}
\usepackage{amsmath}
\usepackage{txfonts}
\usepackage{natbib}
\usepackage{aas_macros}
\usepackage[all]{xy}
\usepackage{booktabs}
\usepackage{multirow}

\usepackage{ifthen}
\ifx\pdftexversion\undefined
\usepackage[dvips]{graphicx}
\else
\usepackage[pdftex]{graphicx}
\usepackage{epstopdf}
\usepackage{thumbpdf}
\fi

\usepackage[all]{hypcap}

\def\del#1{{}}

\newcommand{\dd}{\mathrm{d}}

\title[Weak lensing tomography and CMB lensing]{Parameter constraints from weak lensing tomography of galaxy shapes and cosmic microwave background fluctuations}
\author[Philipp M. Merkel and Bj{\"o}rn Malte Sch\"afer]
{Philipp M. Merkel$^1$\thanks{e-mail: philipp.merkel@urz.uni-heidelberg.de} and Bj{\"o}rn Malte Sch\"afer$^2$\\
${}^1$Institut f{\"u}r Theoretische Astrophysik, Zentrum f{\"u}r Astronomie, Universit{\"a}t Heidelberg, Philosophenweg 12, 69120 Heidelberg, Germany\\
${}^2$Astronomisches Recheninstitut, Zentrum f{\"u}r Astronomie, Universit{\"a}t Heidelberg, Philosophenweg 12, 69120 Heidelberg, Germany}

\begin{document}
\pagerange{\pageref{firstpage}--\pageref{lastpage}}
\pubyear{2017}
\maketitle
\label{firstpage}

\begin{abstract}
Recently, it has been shown that cross-correlating CMB lensing and $3D$~cosmic shear allows to considerably tighten cosmological parameter constraints. We investigate whether similar improvement can be achieved in a conventional tomographic setup. We present Fisher parameter forecasts for a \textit{Euclid}-like galaxy survey in combination with different ongoing and forthcoming CMB experiments. In contrast to a fully three-dimensional analysis we find only marginal improvement. Assuming \textit{Planck}-like CMB data we show that including the full covariance of the combined CMB and cosmic shear data improves the dark energy figure of merit by only three~per~cent. The marginalized error on the sum of neutrino masses is reduced at the same level. For a next generation CMB satellite mission such as \textit{Prism} the predicted improvement of the dark energy figure of merit amounts to approximately $25$~per~cent. Furthermore, we show that the small improvement is contrasted by an increased bias in the dark energy parameters when the intrinsic alignment of galaxies is not correctly accounted for in the full covariance matrix.
\end{abstract}


\section{Introduction}
\label{sec_introduction}
The statistics of cosmic microwave background (CMB) anisotropies, of gravitationally lensed galaxies and the gravitational lensing effect in the CMB are all excellent cosmological probes. CMB fluctuations on the one hand reflect the Universe at its early stages, CMB lensing on the other hand probes cosmic structures at intermediate times, whereas cosmic shear measurements reveal the Universe's late-time evolution. Due to this cosmic complementarity the combination of the various probes is a very powerful tool to constrain cosmological parameters \citep{2002PhRvD..65b3003H}, in particular those concerning dark energy.

Cosmic shear describes the apparent compression and elongation of galaxy shapes due to weak gravitational lensing by foreground matter structures \citep[see][for a review]{2001PhR...340..291B}. The individual deflections are small and not directly accessible to observations but can be inferred statistically from samples of galaxies because the lensing structures are correlated. The power spectrum of the shear components is directly related to the matter power spectrum. Since galaxy shapes are observed on the celestial sphere the shear components are line-of-sight projected quantities. Conventional projection, however, causes an inevitable loss of information because it does not allow to resolve the time evolution of the cosmic density field. For time-resolved cosmic shear measurements photometric redshift information of the background galaxies has to be taken into account. Sufficiently precise redshifts allow to divide the galaxy sample into several redshift bins making the time evolution of the cosmic structures accessible \citep{1999ApJ...522L..21H,2003PhRvL..91n1302J, 2004ApJ...601L...1T,2012MNRAS.423.3445S}. The inter- and intra-bin correlations of tomographic shear measurements then provide a significantly enhanced sensitivity to the equation of state parameter of a potentially time-evolving dark energy component. This sensitivity may be even further enhanced by employing fully three-dimensional methods \citep{2003MNRAS.343.1327H,2005PhRvD..72b3516C,2006MNRAS.373..105H}: $3D$~cosmic shear uses the photometric redshift of each individual galaxy, thereby overcoming the limitations of a coarse redshift binning.

Similarly to the light of distant galaxies photons of the CMB are gravitationally distorted by the intervening large-scale structure \citep[see][for a comprehensive review]{2006PhR...429....1L}. CMB lensing changes the observed CMB power spectra in a characteristic way and the precision of current CMB observations allows to statistically reconstruct the CMB lensing potential from the observed temperature and polarization data \citep{2001PhRvD..64h3005H,2002ApJ...574..566H,2002PhRvD..66f3008O,2014A&A...571A..17P, 2016A&A...594A..15P}. Since all CMB photons emanate from the last scattering surface the estimated CMB lensing potential does not vary with redshift.

Since the intervening large-scale structure lenses CMB radiation as well as the light of galaxies both lensing signals are correlated. This cross-correlation has been successfully detected for a number of different combinations of CMB and weak lensing observations \citep{2015PhRvD..91f2001H, 2015PhRvD..92f3517L, 2016MNRAS.460..434H, 2016MNRAS.459...21K, 2017MNRAS.464.2120S}.
In order to exploit its full information content \citet{2015MNRAS.449.2205K} incorporate the CMB lensing-cosmic shear cross-correlation in the framework of $3D$~cosmic shear. They show that the additional redshift resolved information on the Universe's expansion history and structure growth helps tighten parameter constraints significantly with respect to an analysis discarding this so-called interdatum covariance between CMB lensing and galaxy ellipticities.

Although one might naively think that statistically independent measurements should be best at constraining a model, it is rather the fact that the additional information in the cross-correlation with its parameter constraining power is advantageous to exploit:
Firstly, on large scales there is a non-vanishing correlation of observed galaxy shapes and the CMB temperature due to the integrated Sachs Wolfe (iSW) effect, which has been investigated in a three-dimensional analysis by \citet{2016MNRAS.459.1586Z}, as a generalization of tomographic measurements \citep[e.g.][]{2008PhRvD..78d3519H,2012MNRAS.425.2589J}. Secondly, both weak lensing of the CMB and of galaxies are caused by the very same large scale structure. Consequently, the estimate of the reconstructed CMB lensing potential and the gravitationally lensed galaxy shapes are correlated.

In this work we consider both aspects simultaneously using a tomographic set-up instead of a fully three-dimensional analysis. We compare the parameter constraints obtained in a Fisher matrix forecast by either exploiting or ignoring the interdatum covariance and scrutinize the impact of the coarse binning in redshift on the achievable improvement. Furthermore, we show how the parameter estimation bias, which results from uncorrected systematics, is changed with respect to a conventional analysis when the full covariance matrix of CMB and cosmic shear measurements is included. Specifically, we carry out this exercise for intrinsic alignments as a systematic in weak lensing data. In addition we consider the impact of minute changes in the shape of the matter power spectrum due to baryonic effects. In this, our goal is to consider the entire estimation process with an accurate modeling of parameter inference process for experiments which are being discussed: We present results on a $w$CDM-cosmology with a neutrino-component for \textit{Euclid} and \textit{DES}-like shear experiments in combination with \textit{Planck}, \textit{ACTPol wide} and \textit{Prism}-like CMB observations including the gravitational lensing effect of the CMB.

The structure of this article is as follows: In Section~\ref{sec:formalism} we describe the formalism of our analysis and introduce the different cosmological observables considered in this work along with the different experimental set-ups. The results are presented in Section~\ref{sec:results} and we conclude in Section~\ref{sec:conclusion}.

Throughout this paper we use a spatially flat $\Lambda\mathrm{CDM}$ as reference cosmology. The specific parameter values are $\Omega_\mathrm{m}=0.312$, $\Omega_\gamma=9.167\times 10^{-5}$, $\Omega_\mathrm{b}=0.0483$, $\sigma_8=0.834$, $n_\mathrm{s}=0.9619$, $h=0.67556$, $m_\nu=0.2~\mathrm{eV}$ for the total, radiation and baryonic matter density parameters, respectively, the normalization and spectral index of the power spectrum, the Hubble parameter evaluated today and the sum of massive neutrinos, This choice is compatible with the latest \textit{Planck} data release \citep{2016A&A...594A..13P}.

\section{Formalism}\label{sec:formalism}

\subsection{Cosmological background and structure growth}
The expansion of the homogeneous background in a spatially flat Friedmann-Lema\^itre-Robertson-Walker cosmology is governed by the Hubble function 
\begin{equation}
\frac{H^2(a)}{H^2_0} = 
\frac{\Omega_\gamma}{a^4} + 
\frac{\Omega_\mathrm{m}}{a^3} + 
\left( 1 - \Omega_\mathrm{m}  - \Omega_\gamma\right)
\exp\left( 3\int_a^1 \frac{\dd a'}{a'} \left[ 1 + w_a(a') \right] \right).
\end{equation}
For the time evolution of the dark energy equation of state parameter we adopt the common parametrization \citep{2001IJMPD..10..213C, 2003PhRvL..90i1301L}~$w(a)=w_0 + w_a ( 1 -  a )$, from which we recover our reference cosmology, i.e. a cosmological constant, by setting $(w_0, w_a) = (-1,0)$. 
With the Hubble function it is possible to convert redshift into comoving distance
\begin{equation}
 \chi ( z ) = \int^z_0 \frac{\dd z'}{H(z')}.
\end{equation}

Cosmological structures, i.e. deviations from the homogeneous background, are characterized by the density contrast~$\delta$. 
In the linear regime each of its Fourier modes evolves independently in proportion to the growth function,
i.e. $\delta( \bmath k, a ) = D_+(a)\, \delta_0(\bmath k)$. The growth function is normalized to unity today and fulfills the
following differential equation \citep[e.g.][]{1998ApJ...508..483W,2003MNRAS.346..573L}  
\begin{equation}
\frac{\dd^2}{\dd a^2} D_+(a) + 
\frac{1}{a} \left( 3 + \frac{\dd\ln H}{\dd\ln a} \right) \frac{\dd}{\dd a} D_+(a) - 
\frac{3}{2a^2} \Omega_\mathrm{m} (a) D_+(a) = 0.
\end{equation}
Linear structure growth preserves the statistical properties of the density contrast, which we take to be a statistically homogeneous and isotropic Gaussian random field, completely characterized by its power spectrum
\begin{equation}
 \left\langle \delta (\bmath k ) \delta^* (\bmath k') \right\rangle = (2\upi)^3 \, \delta_\mathrm{D} (\bmath k - \bmath k') \, P_{\delta\delta} ( k ).
\end{equation}
The applicability of linear structure formation ceases on scales smaller than~$\sim 0.1 \, h \, \mathrm{Mpc}^{-1}$ and corrections to the matter power spectrum due to nonlinear clustering need to be taken into account. To this end we employ the halofit approach of \citet{2003MNRAS.341.1311S} along with its extensions proposed by \citet{2012ApJ...761..152T}. We use the implementation provided by the \textsc{cosmic linear anisotropy solving system} \citep[\textsc{class;}][]{2011JCAP...07..034B}. This code is also used to compute CMB temperature and polarization power spectra.

\subsection{Data vector and covariance matrix}
The data vector we are considering comprises tomographic measurements of the weak lensing convergence~$\kappa^i$, an estimate of the CMB lensing potential~$\phi$ and the CMB itself. From the latter we employ both temperature~$\Theta$ and $E$-mode polarization information; its purely lensing induced $B$-mode is exclusively used to reconstruct the CMB lensing potential (cf. Section~\ref{subsubsec:CMB_lensing}). 

We assume that contributions of intrinsically aligned galaxy shapes to the observed weak lensing convergence can be described by a linear alignment model \citep{2001MNRAS.320L...7C}, which is thought to be primarily applicable to elliptical galaxies. In the framework of the linear model changes in galactic shapes reflect the magnitude and orientation of tidal gravitational fields \citep{2004PhRvD..70f3526H,2015SSRv..193....1J}. For disc galaxies more complicated models are required \citep{2001ApJ...559..552C, 2002MNRAS.332..788M}.

Using the harmonic space representation of each observable the data vector reads
\begin{equation}
 \bmath d_{\ell m} = \left(\tilde\kappa_{\ell m}^1, ..., \tilde\kappa_{\ell m}^{N_\mathrm{bin}}, \tilde\phi_{\ell m}, \tilde\Theta_{\ell m}, \tilde E_{\ell m} \right)^\mathrm{t},
\end{equation}
where the superscripted tilde indicates that all sources of observational noise are included, i.e. shape-noise contributions in cosmic shear measurements, instrumental noise in CMB observations and the reconstruction noise of the estimated CMB lensing potential. In case of CMB temperature and polarization it also labels lensing. 

Thus, cosmological information is contained in the covariance of the data vector, which assumes for the case of random Gaussian variables with zero mean the shape
\begin{equation}
 \begin{split}
  \tilde{\textbfss C}(\ell) 
 &\equiv \left\langle \bmath d_{\ell m} \bmath d^+_{\ell' m'} \right\rangle \\
 &\hspace{-0.325cm}=\begin{pmatrix}
  \tilde C^{\tilde\kappa\tilde\kappa}_{11} (\ell) & \cdots & \tilde C^{\tilde\kappa\tilde\kappa}_{1N_\mathrm{bin}} (\ell) & \tilde C^{\tilde\phi\tilde\kappa}_1(\ell) 
  & \tilde C^{\tilde\Theta\tilde\kappa}_1(\ell) & 0\\
   \vdots  & \ddots & \vdots &\vdots & \vdots &\vdots\\
   \tilde C^{\tilde\kappa\tilde\kappa}_{N_\mathrm{bin} 1} (\ell)  & \cdots & \tilde C^{\tilde\kappa\tilde\kappa}_{N_\mathrm{bin}N_\mathrm{bin}} (\ell) & \tilde C^{\tilde\phi\tilde\kappa}_{N_\mathrm{bin}}(\ell)&  \tilde C^{\tilde\Theta\tilde\kappa}_{N_\mathrm{bin}}(\ell)  &0\\
   \vspace{-5pt}
   &\\
   \tilde C^{\tilde\phi\tilde\kappa}_1(\ell) & \cdots & \tilde C^{\tilde\phi\tilde\kappa}_{N_\mathrm{bin}}(\ell) & \tilde C^{\tilde\phi\tilde\phi}(\ell) & \tilde C^{\tilde\phi\tilde\Theta}(\ell) & 0 \\
      \vspace{-5pt} & \\
   \tilde C^{\tilde\Theta\tilde\kappa}_1(\ell) & \cdots & \tilde C^{\tilde\Theta\tilde\kappa}_{N_\mathrm{bin}}(\ell) & \tilde C^{\tilde\phi\tilde\Theta}(\ell) & \tilde C^{\tilde\Theta\tilde\Theta}(\ell)
    & \tilde C^{\tilde\Theta \tilde E}(\ell)\\
       \vspace{-5pt}
    &\\
     0 & \cdots & 0 & 0 & \tilde C^{\tilde \Theta \tilde E}(\ell) & \tilde C^{\tilde E \tilde E}(\ell)
 \end{pmatrix},
 \end{split}
 \label{eq:data_vector_covariance_matrix}
\end{equation}
which is a function of~$\ell$ only due to statistical isotropy, and $\bmath d^+_{\ell m}$ is the Hermitian conjugate of the data vector. Note that we have neglected the small correlation between the CMB $E$-mode polarization and the large-scale structure due to rescattering of the dipole generated by the integrated Sachs-Wolfe-effect \citep{2006JCAP...01..018C}.

\subsection{Signal part of the covariance matrix}
We relegate the noise part of the covariance matrix to the subsequent section and now address the different contributions to its signal part~$\textbfss C(\ell)$. In our analysis, there are three contributions to the observed angular power spectrum of the lensing signal: correlations of gravitationally lensed galaxy shapes $C^{(i,j)}_{GG}(\ell)$, intrinsic shape correlations $C^{(i,j)}_{II}(\ell)$ and the cross-correlation of lensing induced and intrinsically aligned galaxy shapes $C^{(i,j)}_{GI}(\ell)$, respectively:
\begin{equation}
 C_{ij}^{\kappa\kappa}(\ell) = C^{(i,j)}_{GG}(\ell) + C^{(i,j)}_{II}(\ell) + \left(C^{(i,j)}_{GI}(\ell) + C^{(i,j)}_{IG}(\ell)\right) \left( 1 - \frac{1}{2} \delta_{ij}\right).
\end{equation}
For the intrinsic shape correlations we employ the so-called linear alignment model. It assumes that galaxy ellipticities reflect tidal gravitational shear fields sourced by ambient cosmic structures that are in a linear stage of structure formation. But if based on linearly evolving structures \citep{2004PhRvD..70f3526H} \textit{GI}~power on small scales would be underestimated because contributions from nonlinear clustering are not accounted for. We therefore follow \citet{2012MNRAS.424.1647K} and mediate between both regimes by using the geometric mean of the linear and nonlinear matter power spectra in case of the \textit{GI}-term, while the \textit{II}-term is sourced by the linear matter power spectrum only (cf. equation~\ref{eq:differentPowerSpectra}). A competing approach stipulates that the reaction of an elliptical galaxy to a tidal shear field is linear to lowest order and effectively instantaneous, even though the tidal shear field might originate from nonlinearly evolving structures.

The distinction between \textit{GI}- and \textit{IG}-alignments follows from the tomographic setup. They correspond to different geometrical source-lens configurations \citep[cf.][]{2012MNRAS.423.1663T}. Obviously, we have $C^{(i,j)}_{GI}(\ell) = C^{(j,i)}_{IG}(\ell)$.  The mixed terms are negative, indicating that lensing signal and intrinsic shape alignments are anti-correlated \citep[see][for recent reviews]{2015PhR...558....1T,2015SSRv..193...67K}.

Analogously, the cross-correlation of the (reconstructed) CMB lensing defection field (cf. Section~\ref{subsubsec:CMB_lensing}) and the observed cosmic shear field is made up by a lensing and an intrinsic alignment term. We use~$D$ for the reconstructed CMB deflection field in order to unify our notation in the following:
\begin{gather}
 C^{\phi\phi}(\ell) = C^{(0,0)}_{DD}(\ell),\\
  C_{i}^{\phi\kappa}(\ell) = C^{(0,i)}_{DG}(\ell) + C^{(0,i)}_{DI}(\ell),
\end{gather}
which physically describes the alignment of galaxies in structures that are responsible for gravitational lensing of the CMB. Again, the term describing the contributions of intrinsic alignments is negative \citep{2014MNRAS.443L.119H,2014PhRvD..89f3528T,2016MNRAS.461.4343L}.

Finally, we address the cross-correlation between the observed lensing signal and the CMB temperature by focusing on the contributions of the late-time iSW effect \citep[see][for details on the iSW effect]{2014PTEP.2014fB110N}. Here we neglect the contributions of intrinsically aligned galaxy shapes because the iSW effect is prominent only on the largest scales where intrinsic alignments are entirely negligible. Hence, we have
\begin{equation}
 C_{i}^{\Theta\kappa}(\ell) = C^{(0,i)}_{TG}(\ell).
\end{equation}
Moreover, due to the rapid decline of the iSW effect on intermediate and small scales we evaluate its cross correlation with the lensing signal only in the linear regime: In fact, most signal is generated at low multipoles and the small contribution of the nonlinear iSW effect is effectively undetectable because of the high cosmic variance generated by the primary CMB.

The cosmic shear surveys considered in this work (see Section~\ref{subsubsec:cosmicShearSurveys}) cover only parts of the sky making a flat-sky approximation permissible. All observables are line-of-sight projected quantities, which can be related to the density contrast. Their angular power spectra are obtained by means of an appropriate Limber projection \citep{1953ApJ...117..134L} of the three-dimensional power spectrum of the general form: 
\begin{equation}
	C^{(i,j)}_{XY} (\ell) = \int_0^{z_\mathrm{max}} \frac{\dd z}{H(z)}  W^i_X(z) \, W^j_Y(z) \,
		\mathcal{P}_{XY} (k=\ell/\chi(z), z).
\label{eq:angularPowerSpectra}
\end{equation}
The weight function takes on the following values
\begin{equation}
W^i_X ( z ) = 
\begin{cases}
  \displaystyle ( 1 + z ) \int_z^{z_\mathrm{max}} \dd z' \, \dfrac{\chi (z') - \chi (z)}{\chi ( z')} \, n_i(z') 
 & \mbox{if } X = G\\
 \vspace{-10pt}\\ 
\displaystyle\chi^{-1}(z) \, n_i(z) \, H(z)
 & \mbox{if } X = I\\
  \vspace{-10pt}\\
 W^0_D(z) =2 \, ( 1 + z ) \, \dfrac{\chi(z) - \chi(z^\star)}{\chi^2(z)\chi(z^\star)} & \mbox{if } X = D\\
  \vspace{-10pt}\\
  W^0_T(z) =2\, \chi^{-1}(z) \left[ \dfrac{D_+(z)}{(z+1)} + \dfrac{\dd D_+(z)}{\dd z}\right] & \mbox{if } X = T.
\end{cases}
\label{eq:weighting_functions}
\end{equation}
Here we expressed the line-of-sight integration in terms of redshift, where the redshift of (instantaneous) recombination is~$z^\star$. The distribution of observed galaxies in the $i$-th redshift bin is given by~$n_i(z)$. The redshift depth of the cosmic shear survey is limited by~$z_\mathrm{max}$.

The three-dimensional source spectra entering equation~\eqref{eq:angularPowerSpectra} can be expressed in terms of the matter power spectrum. They read
\begin{equation}
 \mathcal P_{Z} (k,z)= \dfrac{9}{4} \Omega^2_\mathrm{m} H^4_0
 \begin{cases}
  P^{\mathrm{nl}}_{\delta\delta}( k,z ) & \mbox{if } Z = GG\\ 
    \vspace{-10pt}\\
     \mathcal{A}^2_I \, P^{\mathrm{lin}}_{\delta\delta}( k ) & \mbox{if } Z = II\\
       \vspace{-10pt}\\
            -\mathcal{A}_I \, \sqrt{P^{\mathrm{lin}}_{\delta\delta}( k )} \dfrac{\sqrt{P^{\mathrm{nl}}_{\delta\delta}(k,z)}}{D_+(z)} & \mbox{if } Z = GI, \, IG\\
                   \vspace{-10pt}\\
                   k^{-4} P^{\mathrm{nl}}_{\delta\delta}( k,z ) & \mbox{if } Z = DD\\ 
                   \vspace{-10pt}\\
                   k^{-2} P^{\mathrm{nl}}_{\delta\delta}( k,z ) & \mbox{if } Z = DG\\
                   \vspace{-10pt}\\
                   -k^{-2}  \mathcal{A}_I \, \sqrt{P^{\mathrm{lin}}_{\delta\delta}( k )} \dfrac{\sqrt{P^{\mathrm{nl}}_{\delta\delta}(k,z)}}{D_+(z)} & \mbox{if } Z = DI\\
                   \vspace{-10pt}\\                   
                    k^{-2} D_+(z)\,P^{\mathrm{lin}}_{\delta\delta}( k ) & \mbox{if } Z = TG.\\
 \end{cases}
 \label{eq:differentPowerSpectra}
\end{equation}
For the amplitude of the intrinsic alignments we choose the commonly adapted parameterization~$\mathcal{A}_I \simeq 8.93 \times 10^{-3} H_0^{-2} \mathcal{I}_\mathrm{A}$ \citep{2004PhRvD..70f3526H, 2007NJPh....9..444B, 2011A&A...527A..26J,2012MNRAS.424.1647K,2013MNRAS.434.1808M,2015MNRAS.449.2205K} and include~$\mathcal{I}_\mathrm{A}$ as free parameter in our analysis. Its fiducial value of  one matches the observational results from low-$z$ SuperCOSMOS data \citep{2002MNRAS.333..501B}.

\begin{figure*}
 \resizebox{\hsize}{!}{\includegraphics[scale=0.75]{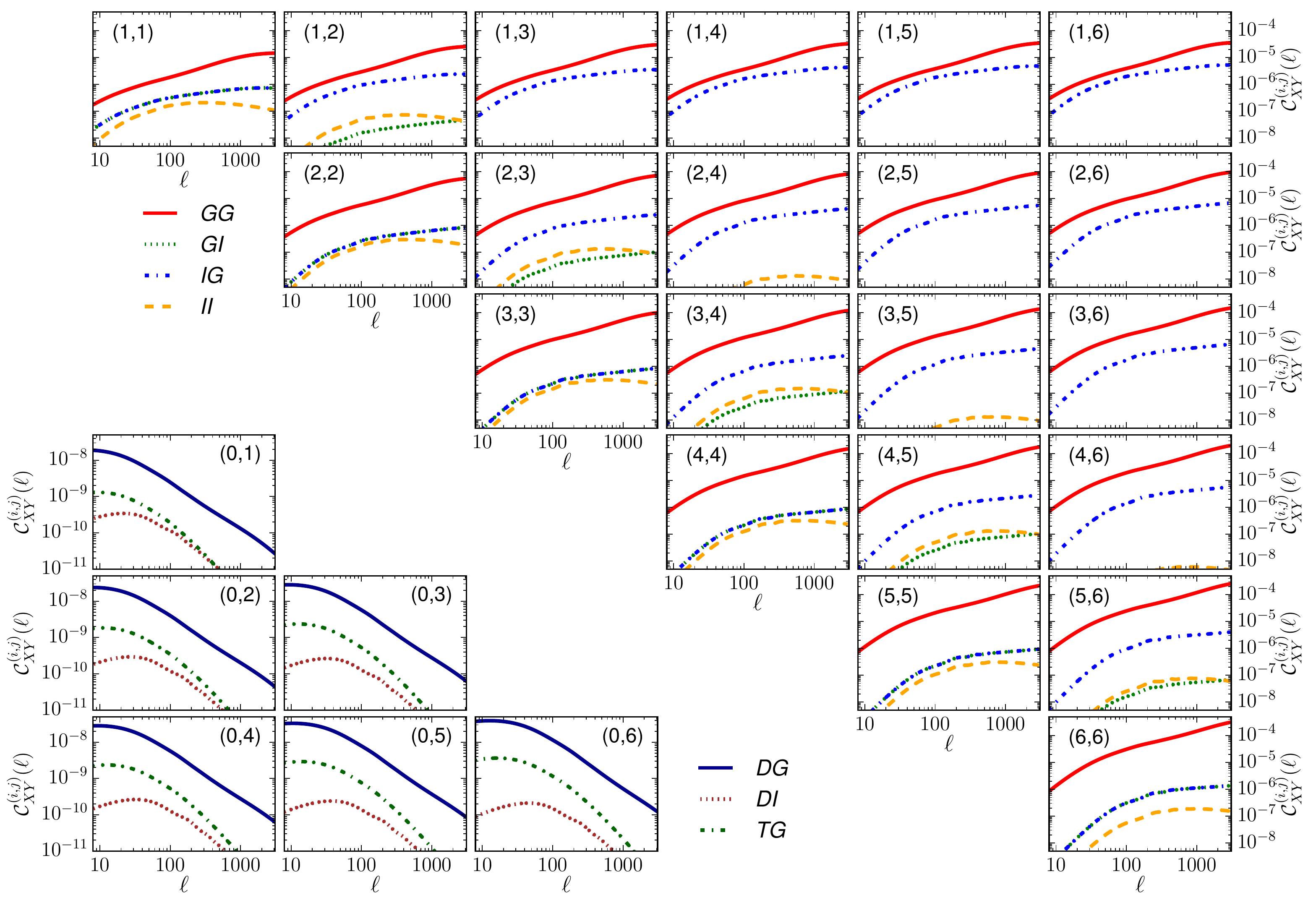}}
 \caption{Power spectra of the weak lensing convergence~(\textit{GG}), intrinsic alignments~(\textit{II}) and their cross-correlation~(\textit{GI}, \textit{IG}) (upper panel); power spectra of the cross-correlation between the CMB lensing potential  and the weak lensing convergence/intrinsic alignments~(\textit{DG}/\textit{DI}) on the one hand and the CMB temperature and the weak lensing convergence~(\textit{TG}) on the other hand (lower panel). The various spectra are derived for a combination of a \textit{Euclid}-like cosmic shear survey using six redshift bins and \textit{Planck}-like CMB observations. The numbers in each subplot indicate the corresponding redshift bins. For convenience we have defined~$\mathcal C^{(i,j)}_{XY}(\ell) \equiv \ell (\ell +1 ) \, \bigl|C^{(i,j)}_{XY}(\ell)\bigr|/(2\upi)$ as a dimensionless representation of the variance without tracking the sign of the correlation.}
 \label{fig:spectra}
\end{figure*}

We illustrate the various power spectra constituting the signal part of the covariance matrix in Figure~\ref{fig:spectra} for a \textit{Euclid}-like weak lensing survey, with six tomographic bins.

\subsection{Surveys and noise properties}\label{subsec:surveys_and_noise_properties}
Having computed the signal part~$\textbfss C(\ell)$ of the data covariance matrix we now proceed with its noise contributions~$\textbfss N(\ell)$. The total covariance matrix is then given by the sum~$\tilde{\textbfss{C}} (\ell) = \textbfss C(\ell) + \textbfss N(\ell)$, where the noise-part is assumed to depend only on the survey characteristics and to be independent from the cosmological model.

\subsubsection{Cosmic shear}\label{subsubsec:cosmicShearSurveys}
Throughout our analysis we assume that the cosmic shear measurements can be carried out with the precision of the forthcoming \textit{Euclid} mission \citep{2013LRR....16....6A}. \textit{Euclid} meets the requirements of a stage~IV cosmic shear experiment. Occasionally, we refer to the \textit{Dark Energy Survey} \citep[\textit{DES,}][]{2005astro.ph.10346T} as a stage~III experiment to illustrate how our results depend on the lensing data quality.

For both surveys the overall distribution of lensed background galaxies is parametrized by
\begin{equation}
 n(z) \propto z^2 \exp \left[  - \left(\frac{z}{z_0}\right)^{\beta} \right].
\end{equation}
It is modulated by a Gaussian with variable width~$\sigma_z = \delta_z (1+z)$ in order to account for photometric redshift errors without catastrophic outliers \citep{2008MNRAS.387..969A}.
Shot noise contributions add to the intrabin power spectra only and are accounted for by a Gaussian ellipticity dispersion~$\sigma_\epsilon$:
\begin{equation}
 N^{(i,j)}_{GG} = \frac{\sigma_\epsilon^2}{\bar n_i} \, \delta_{ij}.
\end{equation}
Here we have introduced the average number of galaxies per bin~$\bar n_i$, which we choose to be identical for all bins.

The specific parameter values for each survey are listed in Table~\ref{tab:cosmicShearSurveyDefinitions}.
\begin{table}
\caption{Cosmic shear survey specifications}
	\begin{tabular}{lccccccc}
		\toprule  
					&\multirow{2}{*}{$\beta$}
					&\multirow{2}{*}{$z_0$} 
					&\multirow{2}{*}{$z_\mathrm{max}$}
					&\multirow{2}{*}{$\delta_z$}
					&\multirow{2}{*}{$f_\mathrm{sky}$}
					&\multirow{2}{*}{$\sigma_\epsilon$}
					&$n_\mathrm{galaxy}$  \\
					&&&&&&&$\bigl[\mathrm{arcmin}^{-2}\bigr]$\\
		\midrule
		\textit{DES} & 1.5 & 0.64 & 2.0 &  0.05  & 0.121& 0.16 &12\\
		\midrule
		\textit{Euclid} & 1.5 & 0.64 & 2.5 &  0.05 & 0.364 &  0.25 & 30 \\
		\bottomrule
	\end{tabular}
\label{tab:cosmicShearSurveyDefinitions}
\end{table}

\subsubsection{CMB}

We characterize the noise contributions to the observed CMB spectra by the (Gaussian) beam size~$\theta_\mathrm{FWHM}$ of the experiment and its instrumental sensitivity with respect to temperature~$\Delta_T$ and polarization~$\Delta_P$, respectively \citep{1995PhRvD..52.4307K}:
\begin{equation}
	\begin{split}
		  N^{\Theta\Theta} (\ell) &= \left( \frac{\Delta_{T}}{T_\mathrm{CMB}} \right)^2 \mathrm{e}^{\ell (\ell +1 ) \theta_\mathrm{FWHM}^2 / 8 \log 2},\\
  N^{EE} (\ell) &= N^{BB} (\ell)= \left( \frac{\Delta_{P}}{T_\mathrm{CMB}} \right)^2 \mathrm{e}^{\ell (\ell +1 ) \theta_\mathrm{FWHM}^2 / 8 \log 2}.	
	\end{split}
\end{equation}
To be specific we summarize the specifications for two ongoing, \textit{Planck} \citep{2006astro.ph..4069T} and \textit{ACTPol wide} \citep{2010SPIE.7741E..1SN}, and one future CMB experiment, \textit{Prism} \citep{2014JCAP...02..006A}, in Table~\ref{tab:cmb_experiments_specifications}. 
\begin{table}
\caption{Specifications of CMB experiments}
\label{tab:cmb_experiments_specifications}
\centering
	\begin{tabular}{lcccc}
\toprule
& $\nu$ & $\theta_\mathrm{FWHM}$ & $\Delta T$ & $\Delta P$ \\
& $ [\mathrm{GHz}]$ & $[\mathrm{arcmin}]$ & $[\mu\mathrm{K\, arcmin}]$ &  $[\mu\mathrm{K\, arcmin}]$ \\
\midrule
\multirow{2}{*}{\textit{Planck}} & 143 & 7.1 & 42.60 & 81.65 \\
	   & 217 & 5.0 & 65.50 & 134.00\\
\midrule
\textit{ACTPol wide} & 150 & 1.4 & 20.00 & 28.00\\
\midrule
\multirow{7}{*}{\textit{Prism}} & 90 & 5.7 & 18.80 & 26.6\\
		    & 105 & 4.8 & 13.80 & 19.60\\
		    & 135 & 3.8 & 9.85 & 13.90\\
		    & 160 & 3.2 & 7.78 & 11.0\\
		    &185  & 2.8 & 7.05 & 9.97\\
		    & 200 & 2.5 & 6.48 & 9.17\\
		    & 220 & 2.3 & 6.26 & 8.85\\
\bottomrule
	\end{tabular}
\end{table}
Whenever there are several frequency bands available we take the inverse weighted sum, i.e.
\begin{equation}
 \frac{1}{N^{XX}(\ell)} = \sum_\nu \frac{1}{N_\nu^{XX} (\ell)},
\end{equation}
as the minimum variance noise for the co-added channels.

\subsubsection{CMB lensing}\label{subsubsec:CMB_lensing}
\begin{figure}
 \resizebox{\hsize}{!}{\includegraphics{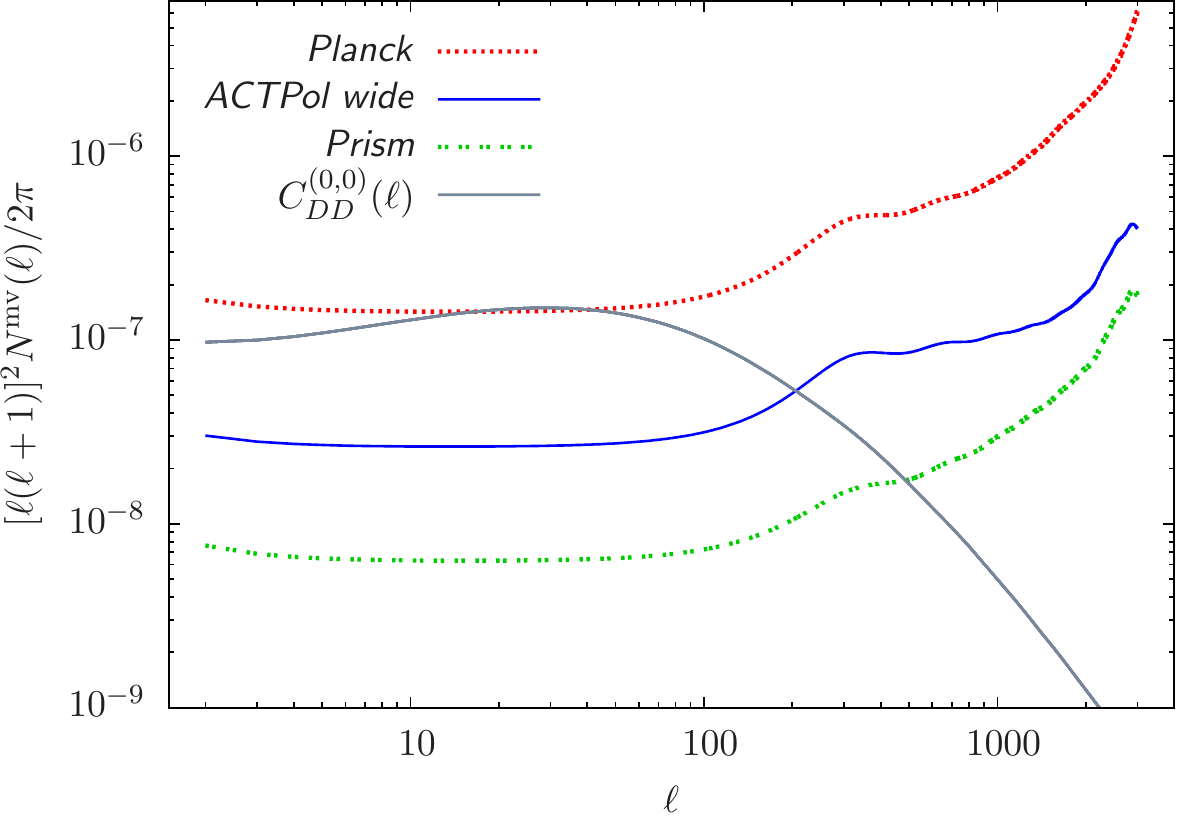}}
 \caption{Reconstruction noise of the CMB lensing potential. For each experiment we show the minimum variance combination of all five estimators that can be constructed from CMB temperature and polarization data.}
 \label{fig:reconstruction_noise}
\end{figure}

The CMB lensing potential is not directly observable but can be statistically reconstructed from the observed, i.e. lensed, CMB temperature and polarization field under the assumption of a statistically homogeneous unlensed CMB. The quadratic estimator proposed by \citet{2002ApJ...574..566H} is optimal in the sense that its Gaussian variance is minimal. This reconstruction noise needs to be added to the lensing potential power spectrum in equation~\eqref{eq:data_vector_covariance_matrix}. Exploiting the lensing induced $B$-mode polarization one can construct five different estimators in total. Their appropriate combination reduces the reconstruction noise even further. We will us this combination and its noise properties in the following and denote it by~$N^\mathrm{mv}(\ell)$. For the explicit expressions of the various estimators, their optimal weighting and the associated reconstruction noise we refer to \citet{2003PhRvD..67h3002O}. In Figure~\ref{fig:reconstruction_noise} we show the reconstruction noise for the three CMB experiments discussed in the previous section. The observed shrinkage in reconstruction noise is mainly due to the improved polarization sensitivity of the next generation CMB missions, making the purely lensing induced $B$-mode accessible. Due to the quadratic structure of the estimator reconstruction noise and cosmic shear signal are uncorrelated at leading order. Thus, we assume
\begin{equation}
	N^{(0,i)}_{DG}(\ell) = N^{(0,i)}_{DI}(\ell) = 0, \quad \forall i
\end{equation}
throughout our analysis but keep in mind that there are higher order corrections for example from the bispectrum of the large-scale structure \citep[][]{2016PhRvD..94d3519B}. Effectively, the reconstruction allows the measurement of the deflection angle spectrum at unit signal to noise-ratio for a range of multipoles even with \textit{Planck}, while \textit{ACTPol wide} and ultimately \textit{Prism} improve on this by half and an entire order of magnitude, respectively.

\subsubsection{ISW effect}
Observations of the iSW effect on the one hand and gravitationally induced shape distortions of galaxies on the other hand have completely uncorrelated noise contributions and we therefore set 
\begin{equation}
 N^{(0,i)}_{TG} (\ell) = 0, \quad \forall i.
\end{equation}

\subsection{Fisher analysis}\label{subsec:FisherAnalysis}

\subsubsection{Fisher matrix}\label{subsubsec:Fisher_matrix}
We invoke the Fisher information matrix \citep{1997ApJ...480...22T} to estimate the statistical accuracy with which cosmological parameters can be constrained from a combined analysis of cosmic shear tomography and the CMB including their various cross-correlations. 
Since, as noted before, all observables have zero mean the Fisher matrix is simply given in terms of the covariance matrix and derivatives thereof with respect to the cosmological parameters~$p_\alpha$,
\begin{equation}
 F_{\alpha\beta} = \frac{1}{2} \, \mathrm{tr} \left( \textbfss C^{-1} \frac{\upartial\textbfss C}{\upartial p_\alpha}
 						\textbfss C^{-1} \frac{\upartial\textbfss C}{\upartial p_\beta}
 					\right).
\end{equation}
Taking the statistical multiplicity of each multipole into account we arrive at
\begin{equation}
 F_{\alpha\beta} = f_\mathrm{sky}\sum_\ell \frac{\left(2\ell + 1\right)}{2} \frac{\upartial C_{ij}(\ell)}{\upartial p_\alpha}
 									\tilde C^{-1}_{jk} (\ell)
 									\frac{\upartial C_{km}(\ell)}{\upartial p_\beta}
									\tilde C^{-1}_{mi} (\ell),
\end{equation}
where summation over repeated indices is implied. The sum over multipoles ranges from~$\ell_{\mathrm{min}}=2$ to~$\ell_\mathrm{max}=3000$ and we assume that all CMB experiments overlap with the fraction of the celestial sphere~$f_\mathrm{sky}$ accessible to the particular cosmic shear survey under consideration. 

In the Fisher matrix framework the likelihood is taken to be Gaussian, which is generally only true for linear models, but which holds approximatively in the vicinity of the peak of tightly constrained likelihoods, because in those cases the nonlinear parameters can be replaced through a Taylor-expansion of the model by effective linear parameters \citep{2017MNRAS.465.4016R}. For pure weak lensing measurements it has been shown that the likelihood is fairly Gaussian \citep{2012JCAP...09..009W} but this does not have to be the case for the extended likelihood in high-dimensional parameter spaces we are considering in this work. Therefore, we would like to emphasize that our results may only serve as an estimate of the attainable accuracy. In particular, they can be used to illustrate the potential improvement which is provided by combining CMB and cosmic shear observations including the interdatum covariance between all quantities. The lower bound on the statistical error of parameter~$p_\alpha$ is then set by the corresponding entry of the inverse Fisher matrix, i.e. $\Delta p_\alpha \geq \sqrt{F^{-1}_{\alpha\alpha}}$, according to the Cram\' er-Rao inequality. This bound includes marginalization over the whole parameter set under consideration. Full knowledge of all parameters but~$p_\alpha$ makes the bound shrink to $\Delta p_\alpha \geq 1/\sqrt{F_{\alpha\alpha}}$.
\begin{table*}
\begin{minipage}{123mm}
\caption{Marginalized 1-$\sigma$~errors predicted for a \textit{Euclid}-like weak lensing experiment combined with CMB data from three different experiments.}
\label{tab:marginalized_predicted_errors}
 \begin{tabular}{lcccccccc}
\toprule
&  & CMB & \multicolumn{2}{c}{{} \hspace{0.25cm} cosmic shear \hspace{0.25cm} {}} &  \multicolumn{2}{c}{cosmic shear $+$ CMB} 
& \multicolumn{2}{c}{cosmic shear $\times$ CMB}\\
\cmidrule(r){4-5}\cmidrule(lr){6-7} \cmidrule(l){8-9}
&   & & \multicolumn{2}{c}{3~bins \hfil 6~bins}  & \multicolumn{2}{c}{3~bins \hfil 6~bins} & 
\multicolumn{2}{c}{3~bins \hfil 6~bins}\\
\midrule
\multirow{10}{*}{\textit{Planck}}&$\Omega_\mathrm{m}$&0.0147&\multicolumn{2}{c}{0.0055\hfil0.0049}&  \multicolumn{2}{c}{0.0046\hfil 0.0042} &    \multicolumn{2}{c}{0.0046\hfil0.0042}\\
&$\sigma_8$&0.0694&\multicolumn{2}{c}{0.007\hfil0.0057}&  \multicolumn{2}{c}{0.0055\hfil 0.0047} &    \multicolumn{2}{c}{0.0054\hfil0.0046}\\
&$\Omega_\mathrm{b}$&0.0017&\multicolumn{2}{c}{0.0114\hfil0.0095}&  \multicolumn{2}{c}{0.0007\hfil 0.0007} &    \multicolumn{2}{c}{0.0007\hfil0.0007}\\
&$m_\nu \, [\mathrm{eV}]$&0.3082&\multicolumn{2}{c}{0.1218\hfil0.0491}&  \multicolumn{2}{c}{0.0448\hfil 0.0338} &    \multicolumn{2}{c}{0.0432\hfil0.0330}\\
&$h$&0.0119&\multicolumn{2}{c}{0.0761\hfil0.0638}&  \multicolumn{2}{c}{0.0046\hfil 0.0044} &    \multicolumn{2}{c}{0.0046\hfil0.0043}\\
&$n_\mathrm{s}$&0.0031&\multicolumn{2}{c}{0.0262\hfil0.0228}&  \multicolumn{2}{c}{0.0025\hfil 0.0025} &    \multicolumn{2}{c}{0.0025\hfil0.0024}\\
&$w_0$&0.6194&\multicolumn{2}{c}{0.0666\hfil0.0553}&  \multicolumn{2}{c}{0.0623\hfil 0.0528} &    \multicolumn{2}{c}{0.0612\hfil0.0523}\\
&$w_a$&1.8812&\multicolumn{2}{c}{0.2304\hfil0.1894}&  \multicolumn{2}{c}{0.1876\hfil 0.1502} &    \multicolumn{2}{c}{0.1846\hfil0.1488}\\
&$\mathcal{I}_\mathrm{A}$& -- &\multicolumn{2}{c}{0.0204\hfil0.0168}&  \multicolumn{2}{c}{0.0194\hfil 0.0159} &    \multicolumn{2}{c}{0.0192\hfil0.0158}\\
\cmidrule{2-9}&FOM&10&\multicolumn{2}{c}{99\hfil139}&  \multicolumn{2}{c}{223\hfil338} &    \multicolumn{2}{c}{234\hfil348}\\
  \hline
  \multirow{10}{*}{\textit{ACTPol wide}}&$\Omega_\mathrm{m}$&0.0054&\multicolumn{2}{c}{\hfil}&  \multicolumn{2}{c}{0.0038\hfil 0.0035} &    \multicolumn{2}{c}{0.0037\hfil0.0034}\\
&$\sigma_8$&0.0196&\multicolumn{2}{c}{\hfil}&  \multicolumn{2}{c}{0.0048\hfil 0.0041} &    \multicolumn{2}{c}{0.0045\hfil0.0039}\\
&$\Omega_\mathrm{b}$&0.0006&\multicolumn{2}{c}{\hfil}&  \multicolumn{2}{c}{0.0005\hfil 0.0005} &    \multicolumn{2}{c}{0.0005\hfil0.0005}\\
&$m_\nu \, [\mathrm{eV}]$&0.0971&\multicolumn{2}{c}{\hfil}&  \multicolumn{2}{c}{0.0393\hfil 0.0312} &    \multicolumn{2}{c}{0.0342\hfil0.0282}\\
&$h$&0.0039&\multicolumn{2}{c}{\hfil}&  \multicolumn{2}{c}{0.0036\hfil 0.0034} &    \multicolumn{2}{c}{0.0036\hfil0.0034}\\
&$n_\mathrm{s}$&0.0021&\multicolumn{2}{c}{\hfil}&  \multicolumn{2}{c}{0.0020\hfil 0.0020} &    \multicolumn{2}{c}{0.0019\hfil0.0019}\\
&$w_0$&0.1690&\multicolumn{2}{c}{\hfil}&  \multicolumn{2}{c}{0.0556\hfil 0.0473} &    \multicolumn{2}{c}{0.0524\hfil0.0453}\\
&$w_a$&0.5129&\multicolumn{2}{c}{\hfil}&  \multicolumn{2}{c}{0.1725\hfil 0.1407} &    \multicolumn{2}{c}{0.1642\hfil0.1359}\\
&$\mathcal{I}_\mathrm{A}$&--&\multicolumn{2}{c}{\hfil}&  \multicolumn{2}{c}{0.0186\hfil 0.0153} &    \multicolumn{2}{c}{0.0179\hfil0.0149}\\
\cmidrule{2-9}&FOM&69&\multicolumn{2}{c}{\hfil}&  \multicolumn{2}{c}{310\hfil 445} &    \multicolumn{2}{c}{364\hfil495}\\
   \hline
   \multirow{10}{*}{\textit{Prism}}&$\Omega_\mathrm{m}$&0.0036&\multicolumn{2}{c}{\hfil}&  \multicolumn{2}{c}{0.0029\hfil 0.0026} &    \multicolumn{2}{c}{0.0027\hfil0.0025}\\
&$\sigma_8$&0.0119&\multicolumn{2}{c}{\hfil}&  \multicolumn{2}{c}{0.0039\hfil 0.0034} &    \multicolumn{2}{c}{0.0032\hfil0.0029}\\
&$\Omega_\mathrm{b}$&0.0003&\multicolumn{2}{c}{\hfil}&  \multicolumn{2}{c}{0.0004\hfil 0.0003} &    \multicolumn{2}{c}{0.0004\hfil0.0003}\\
&$m_\nu \, [\mathrm{eV}]$&0.0697&\multicolumn{2}{c}{\hfil}&  \multicolumn{2}{c}{0.0333\hfil 0.0279} &    \multicolumn{2}{c}{0.0246\hfil0.0218}\\
&$h$&0.0020&\multicolumn{2}{c}{\hfil}&  \multicolumn{2}{c}{0.0025\hfil 0.0024} &    \multicolumn{2}{c}{0.0025\hfil0.0024}\\
&$n_\mathrm{s}$&0.0017&\multicolumn{2}{c}{\hfil}&  \multicolumn{2}{c}{0.0018\hfil 0.0017} &    \multicolumn{2}{c}{0.0017\hfil0.0016}\\
&$w_0$&0.1037&\multicolumn{2}{c}{\hfil}&  \multicolumn{2}{c}{0.0466\hfil 0.0404} &    \multicolumn{2}{c}{0.0412\hfil0.0366}\\
&$w_a$&0.2947&\multicolumn{2}{c}{\hfil}&  \multicolumn{2}{c}{0.1436\hfil 0.1226} &    \multicolumn{2}{c}{0.1326\hfil0.1150}\\
&$\mathcal{I}_\mathrm{A}$& -- &\multicolumn{2}{c}{\hfil}&  \multicolumn{2}{c}{0.0172\hfil 0.0145} &    \multicolumn{2}{c}{0.0164\hfil0.0140}\\
\cmidrule{2-9}&FOM&167&\multicolumn{2}{c}{\hfil}&  \multicolumn{2}{c}{481\hfil 647} &    \multicolumn{2}{c}{671\hfil829}\\
\bottomrule
 \end{tabular}

\medskip
The way how CMB and tomographic cosmic shear data are combined is indicated by a '$+$' and a '$\times$' when the cross-correlation of both data sets is either neglected or accounted for. For comparison we also quote the errors which are attainable when both cosmological probes are considered separately.
\end{minipage}
\end{table*}

The Fisher formalism is also well suited to derive the confidence regions in a two-dimensional parameter subspace. The constraining power of different experimental set-ups then may be assessed against the area of the predicted confidence regions. Therefore a commonly used figure of merit (FOM) is defined by~$\mathrm{FOM}\equiv\sqrt{\det\textbfss F^{(2\times2)}}$, where~$\textbfss F^{(2\times2)}$ indicates the fully marginalized two-dimensional Fisher matrix. Its application to the dark energy parameters~$w_0$ and~$w_a$ is of particular interest \citep{2006astro.ph..9591A}, and we will always refer to the dark energy figure of merit, when using the term FOM.

\subsubsection{Parameter estimation bias}
\label{subsubsec:Bias}
The Fisher matrix approach may be also used to estimate the bias due to systematic effects that are present in the data but not accounted for in the model used in the analysis from which the best-fitting parameter values are derived \citep[e.g.][]{2006MNRAS.366..101H, 2008MNRAS.391..228A}. We assume that the systematics can be described as additive contributions to the expected signal, thus the observed power spectra are given by $C^\mathrm{obs}_\ell = C^\mathrm{mod}_\ell + C^\mathrm{sys}_\ell$. Then, cosmological parameters~$\hat p_\alpha$ that are derived from~$C^\mathrm{obs}_\ell$ while neglecting~$C^\mathrm{sys}_\ell$ differ from the true values~$p^\mathrm{true}_\alpha$ by the bias
\begin{equation}
 b\left[ \hat p_\alpha \right] = \left\langle \hat p_\alpha \right\rangle - \left\langle p^\mathrm{true}_\alpha \right\rangle
 			= F^{-1}_{\alpha\beta} \, B_\beta
 \label{eq:paramEstimationBias}
\end{equation}
with the bias vector
\begin{equation}
 B_\alpha = \frac{1}{2}\mathrm{tr} \left( \textbfss C^{-1}_\mathrm{mod}  \,
 					\frac{\upartial \textbfss C_\mathrm{mod}}{\upartial p_\alpha} \,
					\textbfss C^{-1}_\mathrm{mod} \,
					\textbfss C_\mathrm{sys}
 			\right).
\label{eq:paramEstimationBiasVector}			
\end{equation}
The Fisher matrix which enters equation~\eqref{eq:paramEstimationBias} is built by definition from the covariance matrix of the (incomplete) model, i.e.~$\textbfss C_\mathrm{mod}$.

\section{Results}\label{sec:results}
\begin{figure*}
 \resizebox{\hsize}{!}{\includegraphics[]{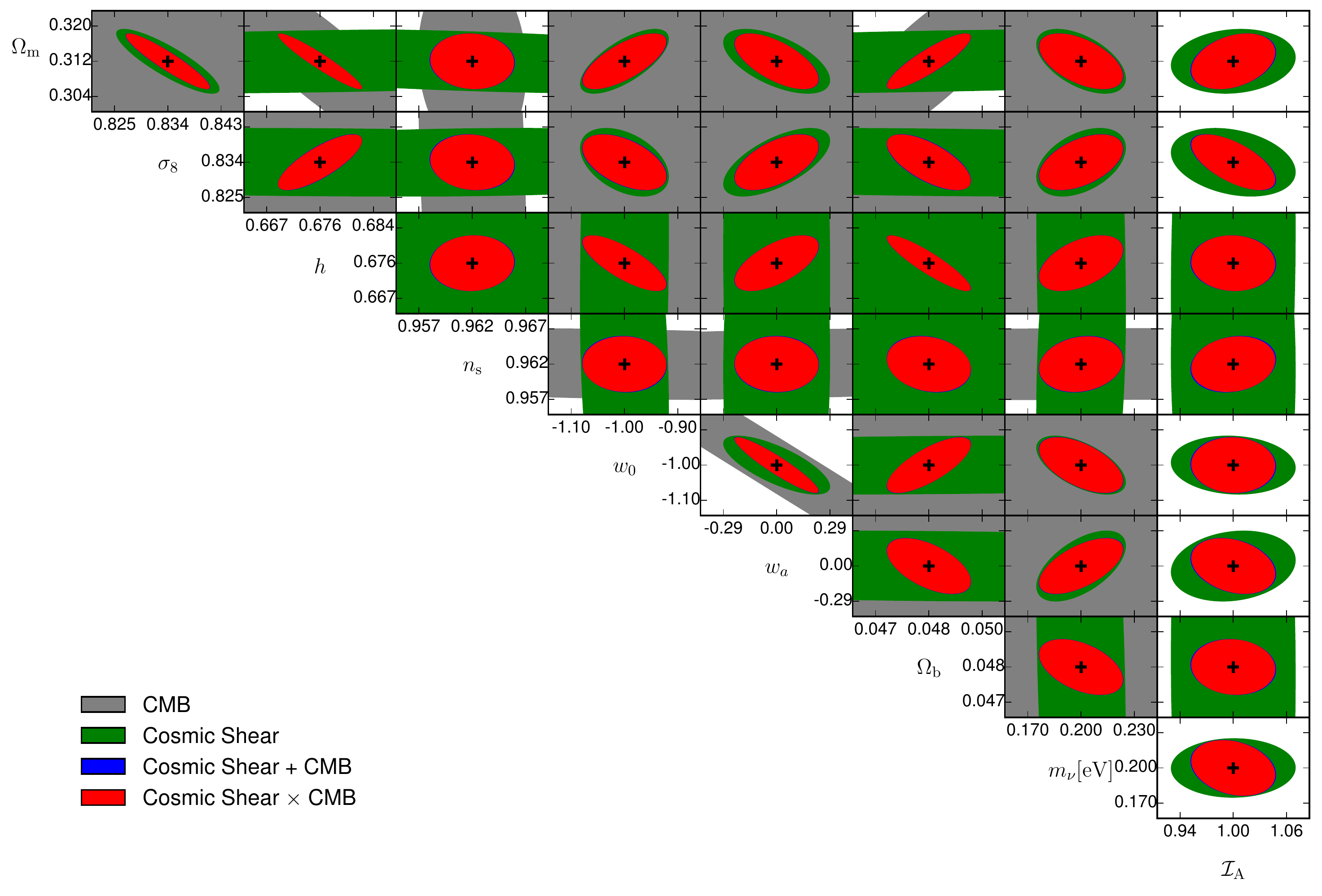}}
 \caption{Forecasted 1-$\sigma$ confidence ellipses for each two-dimensional parameter subspace of the entire nine-dimensional parameter space. The assumed weak lensing survey matches the specifications of \textit{Euclid} and is divided into six redshift bins. It is combined with CMB data from \textit{Planck}, both with primary CMB temperature and polarization spectra as well as the lensing reconstruction. The predicted error bounds include marginalization over the whole parameter set.}
 \label{fig:confidence_regions_all_parameters}
\end{figure*}

We start our analysis with the forecasted errors for a \textit{Euclid}-like shear measurement combined with CMB data from \textit{Planck}. In the upper part of Table~\ref{tab:marginalized_predicted_errors} we compile the marginalized 1-$\sigma$ errors for both experiments individually as well as for their combination. The investigated parameter space is spanned by nine parameters $\Omega_\mathrm{m}$, $\Omega_\mathrm{b}$, $\sigma_8$, $m_\nu$, $h$, $n_\mathrm{s}$, $w_0$, $w_a$ and the intrinsic alignment amplitude $\mathcal{I}_\mathrm{A}$. We consider two different scenarios for combining the data sets: On the one hand-side we treat the two cosmological probes as independent sources of information, on the other hand-side we exploit the full cross-correlation between them. The first case is labeled by~'$+$', whereas~'$\times$' indicates the physically correct usage of cross-correlation information. Weak lensing results are presented for three and six redshift bins, respectively. For the six bin measurement Figure~\ref{fig:confidence_regions_all_parameters} shows the corresponding 1-$\sigma$ confidence ellipses in each two-dimensional parameter subspace.

In general the (uncorrelated) combination of CMB and cosmic shear tightens the parameter constraints significantly due to the complementarity of the observables. Furthermore doubling the number of redshift bins increases the resolution of the Universe's dynamics and thus makes the error bounds of the dark energy parameters shrink. The predicted accuracy of the intrinsic alignment amplitude reaches the per~cent level, indicating the potential power of the two data sets to constrain parameters of more complex alignment models. In the case of the linear alignment model this would imply for example tests of the virial equilibrium that determines the shape of a system under the influence of a perturbed potential.

However, adding cross-correlation information hardly improves on these constraints. The variation is only at the per cent level. Correspondingly, the shrinkage of the two-dimensional confidence region in the dark energy parameter plane is rather modest; the FOM increases by roughly three per cent. The error of the sum of neutrino masses is effectively unchanged. These findings are highly contrasting with the results for a fully three-dimensional analysis. \citet{2015MNRAS.449.2205K} showed that in the framework of $3D$~cosmic shear improvement of the dark energy parameters amounts up to 30 per cent when the full covariance of \textit{Euclid}- and \textit{Planck}-like data is taken into account. Also, the predicted (marginalized 1-$\sigma$) error of the neutrino mass improves by about 25 per cent and the intrinsic alignment parameter may even be measured with almost doubled accuracy. 

One possible explanation for this discrepancy can be seen in the power of $3D$~comic shear to resolve the Universe's dynamics on the one hand and the limited redshift-resolution of lensing tomography on the other hand. Expansion history and structure growth are highly sensitive to dark energy and the neutrino mass. However it is surprising that the resolution of even six redshift bins would be too coarse to exploit the information conveyed by the CMB lensing-cosmic shear cross-correlation. We therefore believe that it is more likely that the observed discrepancy is caused by the additional systematic effects accounted for in the analysis of \citet{2015MNRAS.449.2205K}. In addition to intrinsic alignments of galaxies \citet{2015MNRAS.449.2205K} consider a multiplicative bias due to galaxy shape measurement systematics which is not included in our analysis. Combining cosmic shear observations, which are affected by these systematics, with CMB observations, which are not, helps constrain the multiplicative bias. From this improvement then other parameters may benefit, too. Following this line of argument the gain in parameter constraints found by \citet{2015MNRAS.449.2205K} reflects the fact that their results derived without CMB lensing-cosmic shear cross-correlation information are more uncertain than those obtained including the interdatum covariance because of the possibility of multiplicative biases, which would affect, for example, conclusions on growth rates. Since multiplicative biases are not accounted for in our modeling we do not observe a gain similar to the results of \citet{2015MNRAS.449.2205K}.
\begin{figure*}
\includegraphics[scale=0.4125]{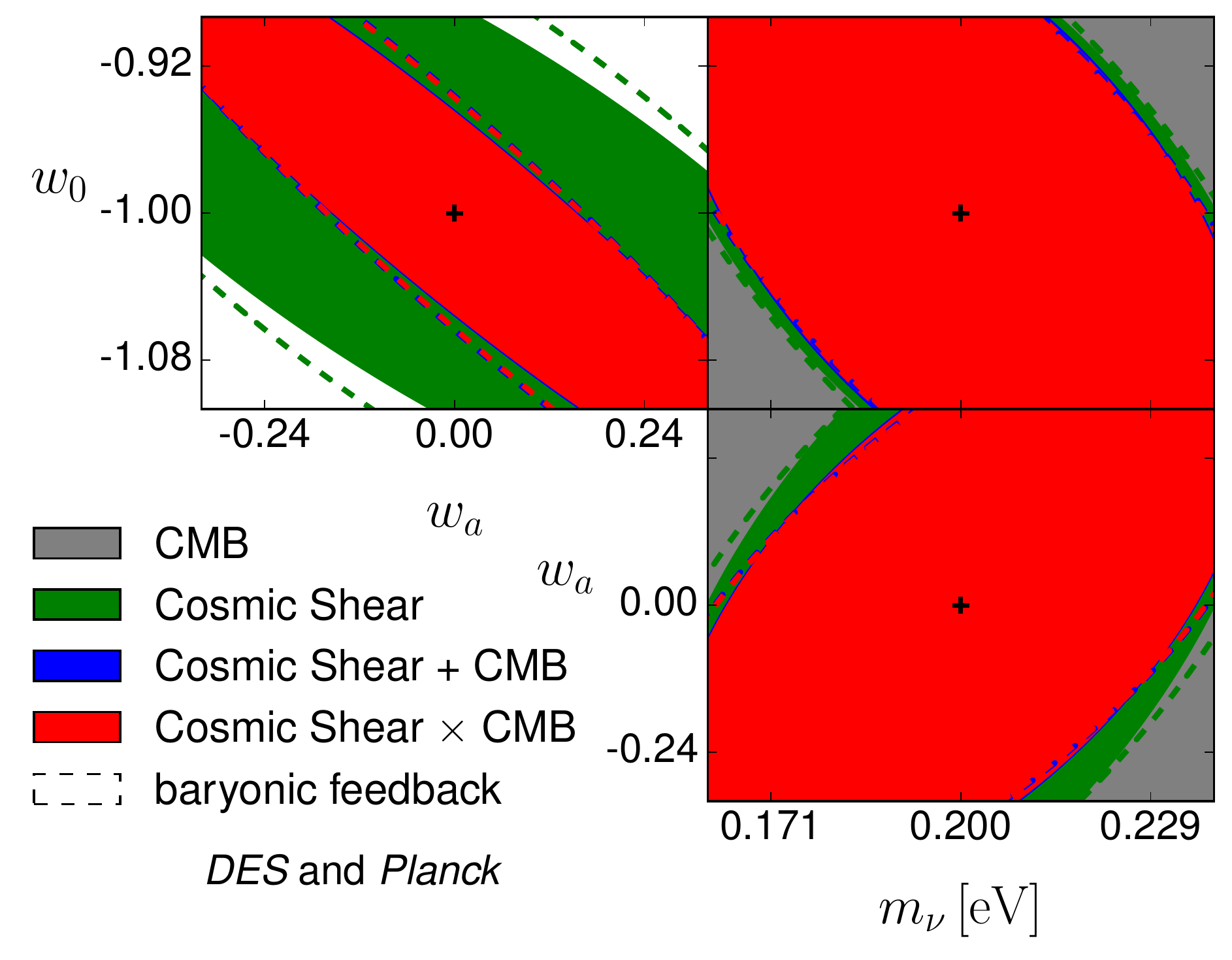}\hfil%
\includegraphics[scale=0.4125]{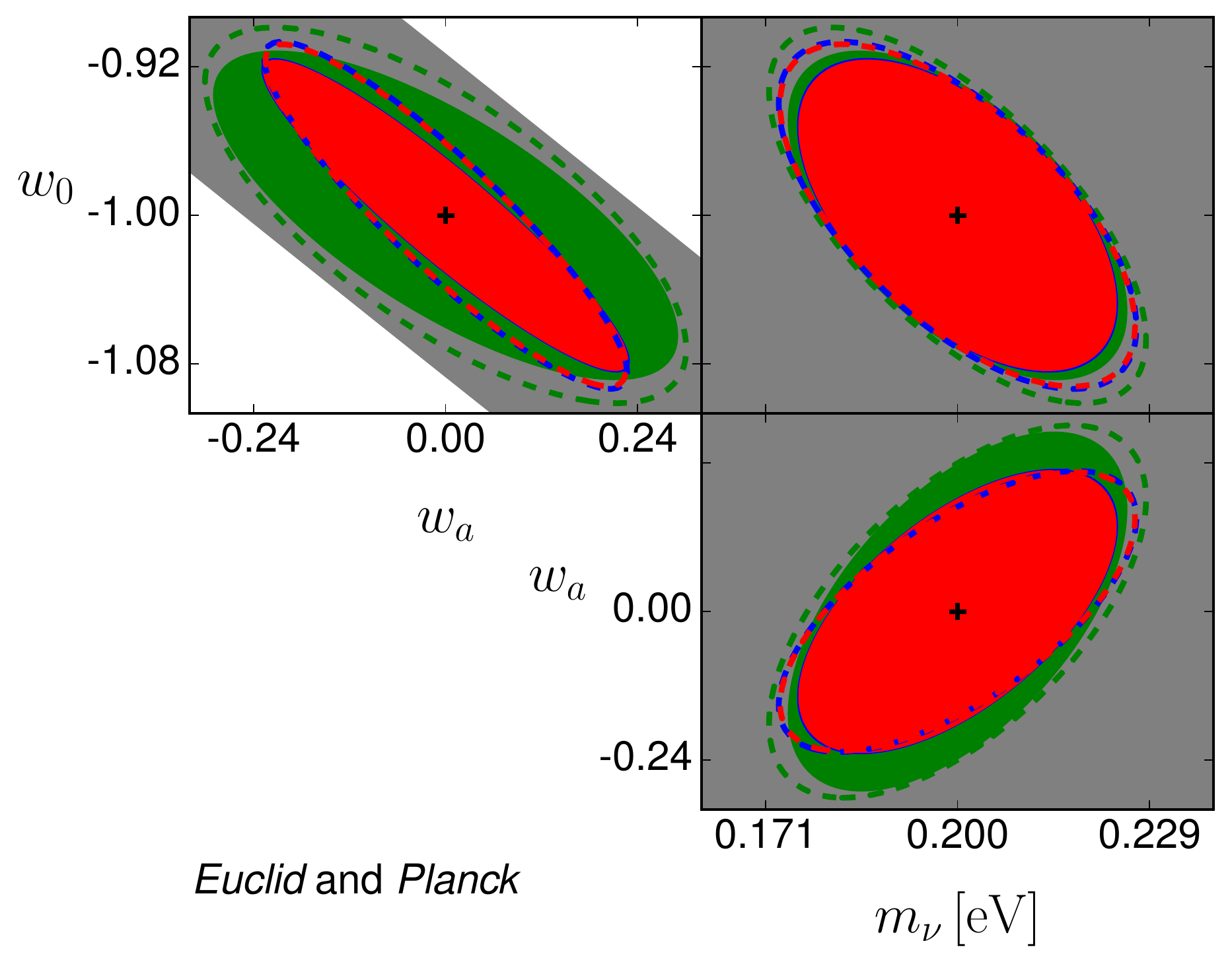}\\%
\includegraphics[scale=0.4125]{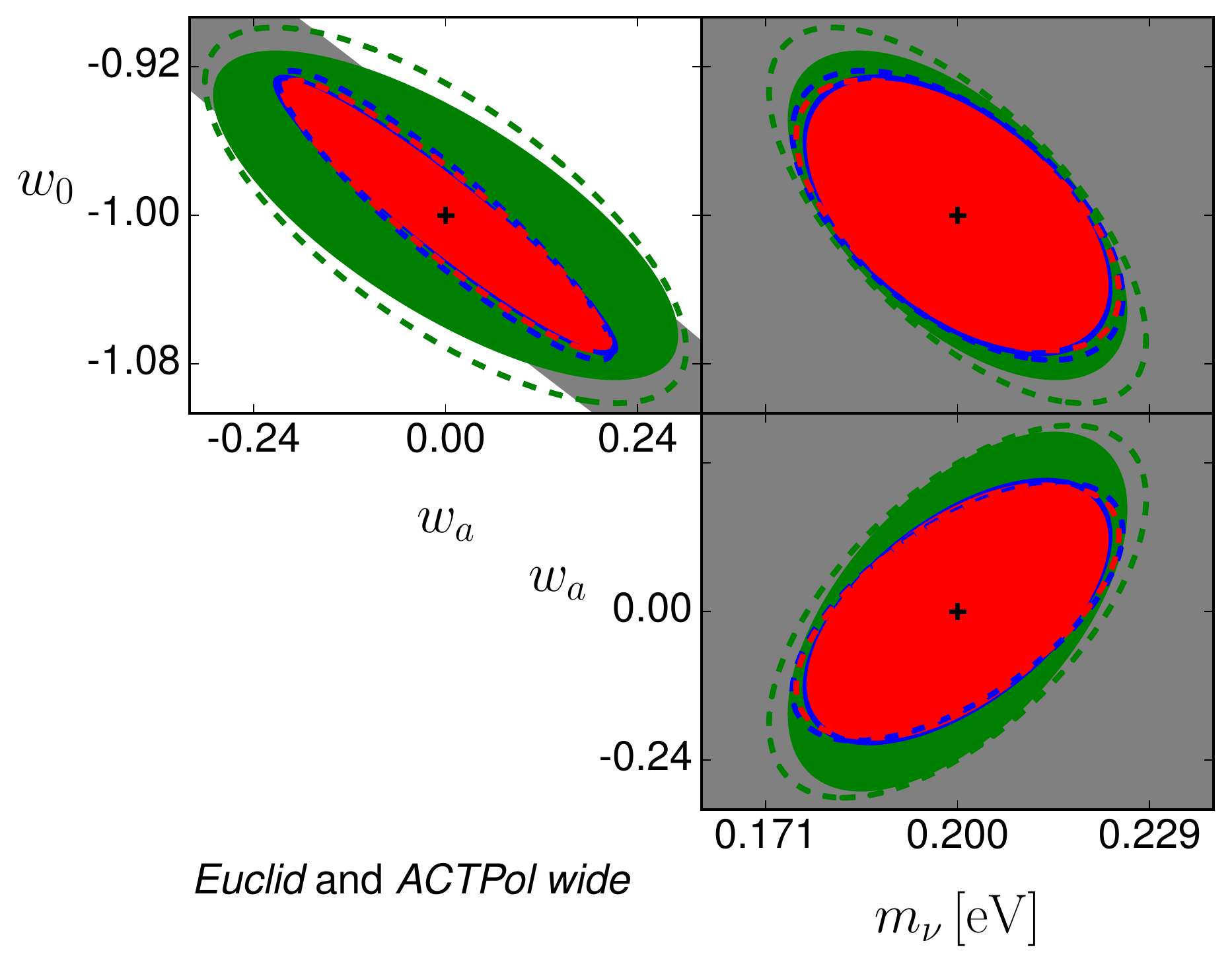}\hfil%
\includegraphics[scale=0.4125]{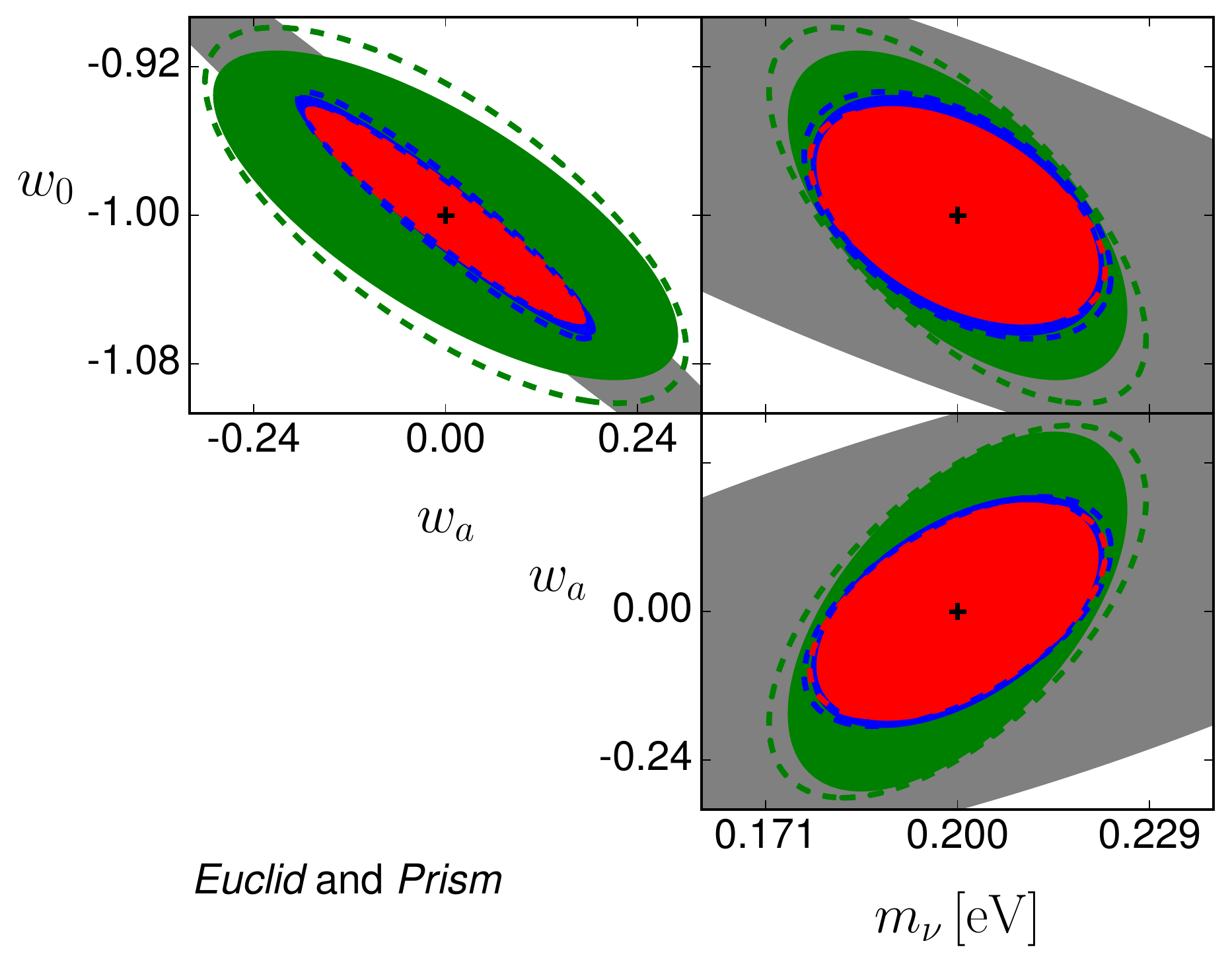}%
 \caption{Predicted 1-$\sigma$ confidence regions in the two-dimensional parameter subspaces spanned by the parameters describing the dark energy equation of state evolution and the sum of neutrino masses. Forecasts are shown for \textit{Euclid} and \textit{DES}-like six bin tomographic weak lensing measurements in combination with three different CMB experiments.}
 \label{fig:dark_energy_confidence_ellipses}
\end{figure*}

In a next step we investigate how the forecasted errors change when higher quality CMB data becomes available. Browsing through Table~\ref{tab:marginalized_predicted_errors}, we find that the differences between measurements, which ignore/include cross-correlation information, become more prominent as CMB data improves, but the differences are still small. Including the CMB lensing-cosmic shear cross-correlation enlarges the dark energy FOM by 10 and 25 per cent for \textit{ACTPol~wide} and \textit{Prism}, respectively. It is interesting to note that although the cross-correlation itself does not dependent on the experimental noise level it helps narrow parameter constraints more efficiently when combined with an improved estimate of the CMB lensing potential.

In Figure~\ref{fig:dark_energy_confidence_ellipses} we visualize the results for the different CMB experiments focusing on  the 1-$\sigma$ confidence regions in the two-dimensional parameter spaces spanned by the dark energy equation of state parameters and the sum of neutrino masses. Additionally, we show the forecasts obtained by combining \textit{DES} data with \textit{Planck} CMB observations to contrast the stage~III with the stage~IV shear experiment. In case of the stage~III shear experiment the parameter constraints do not benefit at all from including CMB lensing-cosmic shear cross-correlation information.

The results presented so far have been derived using the revised halofit model to describe the nonlinear contributions to the matter power spectrum. Nonlinear structure growth boosts the small scale power considerably and therefore helps tighten weak lensing constraints. Especially the nearest redshift bin receives contributions from very small scales, i.e. very large wave numbers~$k_\mathrm{max} \sim 1 \, h \, \mathrm{Mpc}^{-1}$. However, it has been found in numerical simulations that baryonic feedback wipes out small scale fluctuations thereby affecting weak lensing measurements \citep{2011MNRAS.417.2020S,2013MNRAS.434..148S}. Since the CMB lensing signal originates from larger scales it is expected to be less sensitive to baryonic feedback than cosmic shear. We investigate how the parameter constraints derived with and without including the cross-correlation in the two data sets may vary in the presence of baryonic feedback. We adapt the ansatz of \citet{2011MNRAS.417.2020S} to describe the suppression of small-scale power due to baryonic effects: 
\begin{equation}
 P_\mathrm{bf} (k,z) = \left[ A_\mathrm{S} \, \left( \frac{P_\mathrm{tot}  (k)}{P_\mathrm{CDM}(k)}- 1 \right) + 1 \right]  P^\mathrm{nl}_{\delta\delta} (k, z). 
\end{equation}
The ratio of the total matter to dark matter power spectrum is assumed to be redshift-independent on all relevant scales and is one for scales larger than~$0.1 \, h \,\mathrm{Mpc}^{-1}$.

In order to make a possible influence of baryonic feedback clearly identifiable in our analysis we set the amplitude of suppression~$A_\mathrm{S}$ to 3, while a realistic range is~$A_\mathrm{S} \sim 0 \ldots 1$.
Looking at Figure~\ref{fig:dark_energy_confidence_ellipses} we see the expected enlargement of the error ellipses due to the reduced power on small scales. However, the increase of the confidence regions in the presence of baryonic feedback is almost identically for all shown ellipses and hence, there is no substantial gain when adding cross-correlation information.

Finally, we consider the parameter estimation bias which results from erroneously interpreting the observed galaxy shapes without accounting for intrinsic alignments. For simplicity we assume that intrinsic alignments are fully described by the linear alignment model in the form of equation~\eqref{eq:angularPowerSpectra}. Thus, in terms of Section~\ref{subsubsec:Bias} the power spectra~$C^{(i,j)}_{GG}(\ell)$, $C^{(0,i)}_{DG}(\ell)$ and~$C^{(0,i)}_{TG}(\ell)$ enter the covariance matrix $\textbfss C^\mathrm{mod}_\ell$, while $\textbfss C^\mathrm{sys}_\ell$ exclusively comprises~$C^{(i,j)}_{II}(\ell)$, $C^{(i,j)}_{GI}(\ell)$, $C^{(i,j)}_{IG}(\ell)$ and~$C^{(0,i)}_{DI}(\ell)$.

We would like to recall that there are also other alignment models \citep{2001ApJ...559..552C, 2002MNRAS.332..788M} and emphasize that the results may depend on the chosen parametrization \citep{2007NJPh....9..444B, 2012MNRAS.424.1647K, 2013MNRAS.435..194C,2015arXiv150607366S}. In particular, the long-ranged \textit{GI}-alignments, absent in quadratic alignment models for Gaussian fluctuations, are expected to substantially contribute to the bias on intermediate scales with a negative sign, while contributions of the short-ranged \textit{II}-alignments are expected to be small (cf. Figure~\ref{fig:spectra}). Besides their dependence on the choice of alignment model it is important to keep in mind that the results derived using equation~\eqref{eq:paramEstimationBias} may serve only as a qualitative indicator of how strong the results are biased. The adapted formalism is not suited for a quantitative analysis once the bias exceeds~$\sim 1-2\sigma$. Thus, biases of several~$\sigma$ must not be taken literally but rather reveal that the derived parameter constraint is significantly biased if the systematic effect, i.e. intrinsic galaxy shape correlations, are not accounted for.

\begin{figure}
 \resizebox{\hsize}{!}{\includegraphics{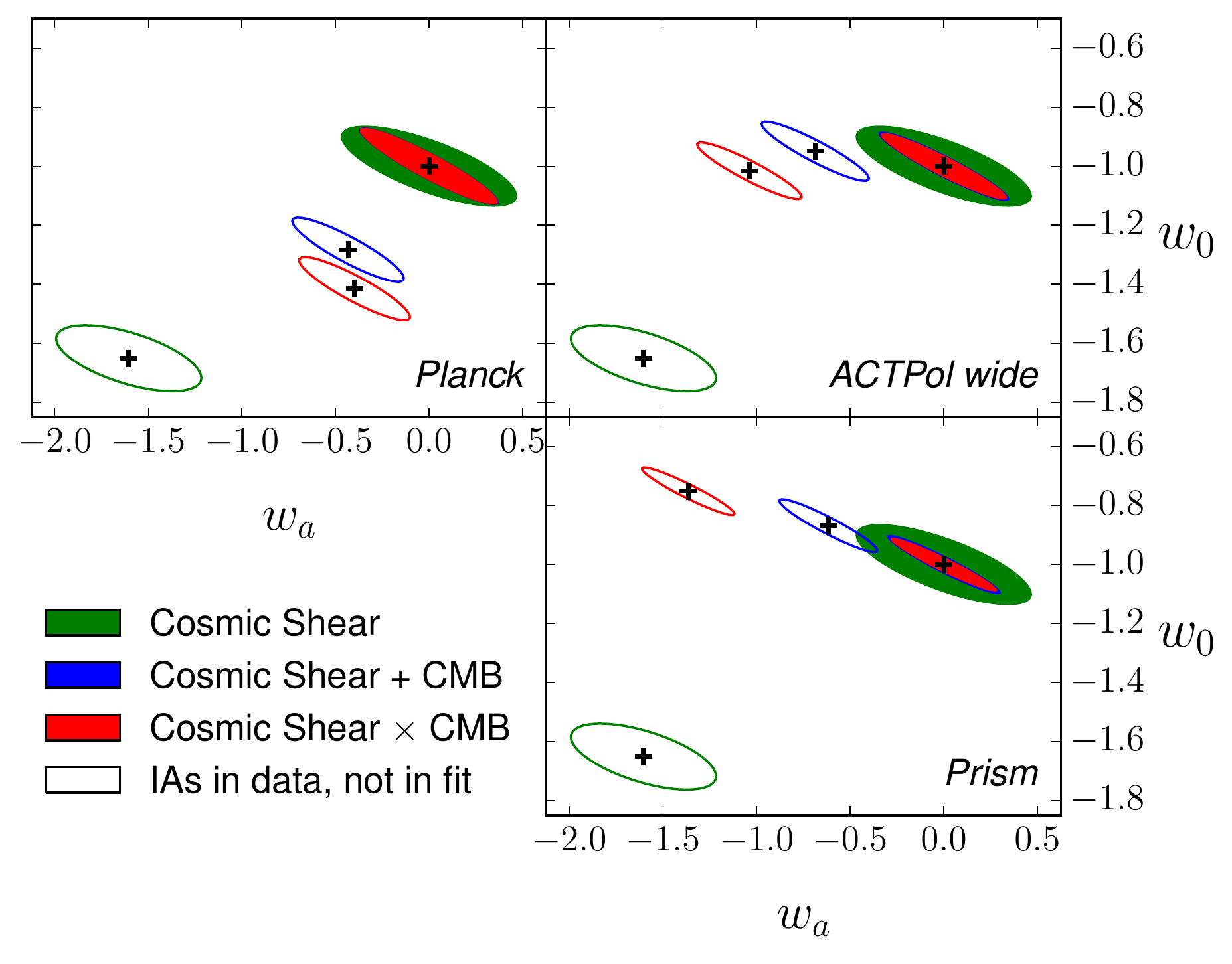}}
 \caption{Parameter estimation biases in the dark energy parameters~$w_0$ and~$w_a$ for a \textit{Euclid}-like cosmic shear survey in combination with different CMB experiments.}
 \label{fig:bias}
\end{figure}

\begin{figure}
 \resizebox{\hsize}{!}{\includegraphics{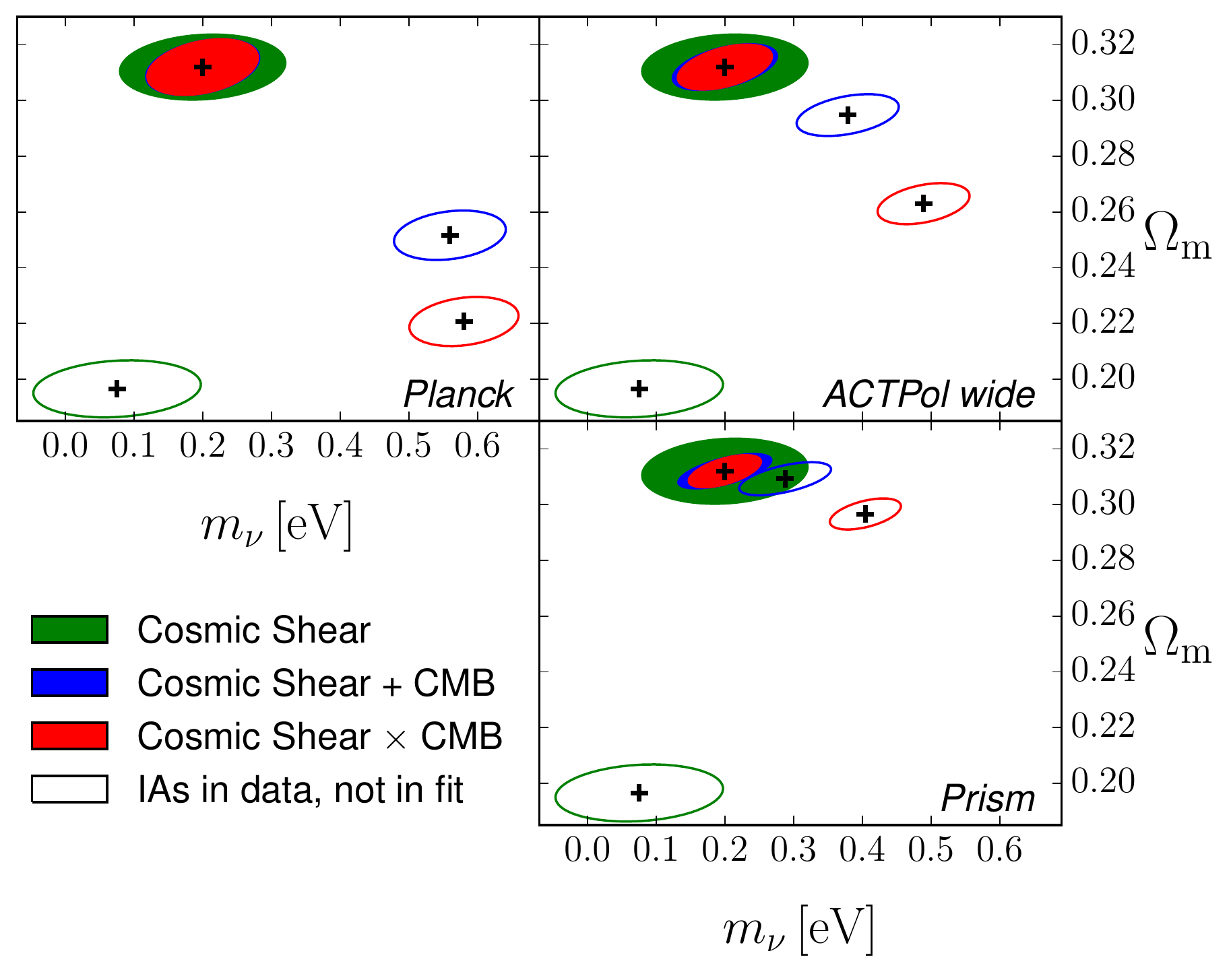}}
 \caption{Parameter estimation biases in the matter density $\Omega_\mathrm{m}$ and the sum of massive neutrinos $m_\nu$ derived when combining \textit{Euclid}-like weak lensing observations with different CMB experiments.}
 \label{fig:biasOmegaMMNeutrino}
\end{figure}

In Figure~\ref{fig:bias} we compile the 1-$\sigma$ confidence limits on the dark energy parameters obtained from \textit{Euclid}-like lensing data in combination with the three different CMB experiments considered in this work. We distinguish the following scenarios: cosmic shear data alone $(i)$, cosmic shear and CMB data in combination with $(ii)$ and without $(iii)$ their cross-correlation. For all three scenarios we show the results obtained when intrinsically aligned galaxies are present in the data and either accounted for (filled ellipses) or erroneously neglected (open ellipses). Quantifying the bias by the distance between the fiducial parameter value and the centre of the contour which ignores intrinsic alignments we find that the bias is largest in case that no CMB data are included. Conversely, it is smallest in case that CMB and cosmic shear data are treated as independent cosmological probes. The ellipses are differently shifted in the $w_0$-$w_a$-plane for the various CMB experiments. However, the bias increases with improving CMB data when the correlation of both data sets is included: almost overlapping for a \textit{Planck}-like CMB survey the (biased) 1-$\sigma$ confidence regions move more and more apart for the anticipated CMB observations by \textit{ACTPol~wide} and \textit{Prism}, respectively. In case of \textit{Prism} the biased 1-$\sigma$ confidence region, which has been obtained ignoring cross-correlation information, and the fiducial ellipse derived from cosmic shear data alone finally intersect. This is in agreement with our previous finding that the better the available CMB data the larger the influence of cross-correlation information on the derived parameter constraints. Similar results are found for the matter density and the sum of massive neutrinos as shown in Figure~\ref{fig:biasOmegaMMNeutrino}.

\section{Conclusion}\label{sec:conclusion}
The subject of this paper is an investigation of the parameter-constraining power of the primordial CMB temperature and polarization anisotropies, the gravitational lensing effect on the CMB and weak lensing of galaxy shapes with particular emphasis on three aspects: $(i)$ We take the full covariance into account and construct all cross-spectra, most notably correlations between the two lensing effects, the cross-correlation between both lensing effects and the intrinsic shapes of galaxies, and cross-correlations between large-scale structure tracers and the CMB generated by the iSW effect. $(ii)$ We investigate how the parameter inference process reacts to the inclusion of systematics, firstly of corrections to the matter spectrum due to baryonic processes and secondly, due to intrinsic shape correlations of galaxies as determined by a linear tidal shearing model. $(iii)$ We carry out the forecasting exercise for a range of current and future experiments, both on the side of the CMB and on the side of cosmic shear, in order to follow how the total signal improves and how constraints tighten, but also to understand how the systematic errors increase and to what limits they need to be controlled. As a baseline cosmological model family, we use a $w$CDM-type cosmology with the baryon density $\Omega_\mathrm{b}$ and the sum of neutrino masses $m_\nu$ as additional fundamental parameters, and describe the intrinsic alignment model with a single parameter which is unconstrained by any prior information.
\begin{enumerate}
\item{The precision results for the baseline $w$CDM-cosmology shows clearly a hierarchy in the cosmological parameters, where~$\Omega_\mathrm{m}$, $\Omega_\mathrm{b}$ and~$\sigma_8$ are the best constrained parameters with forecasted precisions on the $10^{-3}$-level, and where~$h$ and~$n_\mathrm{s}$ benefit from improving the CMB data. The dark energy equation of state parameters as well as the neutrino masses get boosts in their precision by the inclusion of large-scale structure data in addition to CMB data, reaching per~cent levels on~$w_0$ and between ten and twenty per~cent on~$w_a$, based on CMB and weak lensing data alone. Moving from \textit{DES} to \textit{Euclid} in this combination of probes reduces errors by roughly a factor of two on dark energy properties and neutrino masses.}
\item{The amplitude of intrinsic alignments (from a linear alignment model) is constrained by its signal in the lensing survey as well as by the cross-correlation with the CMB lensing signal with an error of a few per~cent on a parameter that is known from current observations to be of order unity. This implies, that investigations of alignment models by this combination is in fact possible without infringing too seriously on the statistical precision on the fundamental cosmological parameters. A measurement of~$\mathcal{I}_\mathrm{A}$ would allow to investigate the dynamical state of dark matter haloes from the way they react to tidal gravitational fields by changing their apparent shape.}
\item{If ignored, intrinsic alignments would impact on the parameter estimation process: The dark energy equation of state parameters~$w_0$ and~$w_a$ would both be significantly biased towards negative values based on weak lensing data alone. High quality CMB data would remedy this by removing much of the bias in~$w_0$. This would imply that a $\Lambda$CDM-cosmology could be mistaken for a cosmology with a strongly varying equation of state. Perhaps even more dramatic are estimation biases in the neutrino masses, which are measured to be too small due to the influence of intrinsic alignments.}
\item{Baryonic corrections to the matter power spectrum \citep{2011MNRAS.417.2020S,2013MNRAS.434..148S} influence the magnitude of statistical errors equally for both types of analysis, i.e. with and without inclusion of CMB lensing-cosmic shear cross-correlation information.}
\item{Concerning the forecasted statistical precision we are less optimistic than \citet{2015MNRAS.449.2205K} who differ from our analysis by using $3D$~weak lensing instead of lensing tomography. While $3D$~cosmic shear is able to capture more of the cosmological information because of the absence of radial binning we do not reproduce the considerable improvement of statistical errors when including the interdatum covariance: 
Whether even six-bin tomography is lagging much behind $3D$~weak lensing given the additional data from the primordial CMB and from CMB lensing, or whether a difference in data modelling \citep[in contrast to][we do not account for multiplicative biases due to cosmic shear measurement systematics]{2015MNRAS.449.2205K} is responsible, is difficult to say. At least, in a preceding study \citep{2016MNRAS.459.1586Z}, where we combined $3D$~weak lensing as a large-scale structure tracer for isolating the iSW effect, we could only find comparatively small improvements on the statistical precision. Furthermore, if systematic effects like intrinsically aligned galaxies are not correctly accounted for in the analysis the results derived using the interdatum covariance are more biased than those obtained by a conventional analysis ignoring the cross-correlation of CMB lensing and galaxy lensing.}
\item{A complete description of the interdatum covariance is essential for statistical inference, and we believe to have included all leading terms at the level of two-point correlations of both the CMB and the weak lensing data: These include the \textit{GI}-correlations between intrinsic alignments and weak lensing, the correlation between galaxy shapes and the CMB deflection field, and the iSW effect generated by the large-scale structure.}
\end{enumerate}

In summary, we would like to emphasize that in particular for dark energy per~cent-level accuracy on~$w_0$ and~$w_a$ is within reach based exclusively on CMB and cosmic shear data.

\section*{Acknowledgements}
We would like to thank the anonymous referee for valuable comments which helped improve our presentation.

\bibliography{references}
\bibliographystyle{mnras}


\bsp

\label{lastpage}

\end{document}